\documentclass[journal=jacsat]{achemso}
\usepackage{chemformula}
\usepackage[T1]{fontenc}
\usepackage{wasysym}
\usepackage{mathtools}
\usepackage{siunitx}
\usepackage{lineno}
\usepackage[normalem]{ulem}
\DeclarePairedDelimiterXPP\BigOSI[2]%
  {\mathcal{O}}{(}{)}{}%
  {\SI{#1}{#2}}

\author{Aleksandr Bashkatov}
\affiliation[HZDR]
{Institute of Fluid Dynamics, Helmholtz-Zentrum Dresden-Rossendorf,\\Bautzner Landstrasse 400, 01328 Dresden, Germany}
\alsoaffiliation[UT]
{Physics of Fluids Group, Max Planck Center for Complex Fluid Dynamics,\\MESA+ Institute and J. M. Burgers Centre for Fluid Dynamics, University of Twente,\\P.O. Box 217, 7500AE Enschede, Netherlands}
\email{a.bashkatov@hzdr.de}

\author{Florian B{\"u}rkle}
\affiliation[TUD2]
{Laboratory for Measurement and Sensor System Techniques,\\Faculty of Electrical and Computer Engineering, Technische Universit{\"a}t Dresden,\\Helmholtzstr. 18, 01069 Dresden, Germany}

\author{Çayan Demirkır}
\affiliation[UT]
{Physics of Fluids Group, Max Planck Center for Complex Fluid Dynamics,\\MESA+ Institute and J. M. Burgers Centre for Fluid Dynamics, University of Twente,\\P.O. Box 217, 7500AE Enschede, Netherlands}

\author{Wei Ding}
\affiliation[HZDR]
{Institute of Fluid Dynamics, Helmholtz-Zentrum Dresden-Rossendorf,\\Bautzner Landstrasse 400, 01328 Dresden, Germany}

\author{Vatsal Sanjay}
\affiliation[UT]
{Physics of Fluids Group, Max Planck Center for Complex Fluid Dynamics,\\MESA+ Institute and J. M. Burgers Centre for Fluid Dynamics, University of Twente,\\P.O. Box 217, 7500AE Enschede, Netherlands}

\author{Alexander Babich}
\affiliation[HZDR]
{Institute of Fluid Dynamics, Helmholtz-Zentrum Dresden-Rossendorf,\\Bautzner Landstrasse 400, 01328 Dresden, Germany}

\author{Xuegeng Yang}
\affiliation[HZDR]
{Institute of Fluid Dynamics, Helmholtz-Zentrum Dresden-Rossendorf,\\Bautzner Landstrasse 400, 01328 Dresden, Germany}

\author{Gerd Mutschke}
\affiliation[HZDR]
{Institute of Fluid Dynamics, Helmholtz-Zentrum Dresden-Rossendorf,\\Bautzner Landstrasse 400, 01328 Dresden, Germany}

\author{J{\"u}rgen Czarske}
\affiliation[TUD2]
{Laboratory for Measurement and Sensor System Techniques,\\Faculty of Electrical and Computer Engineering, Technische Universit{\"a}t Dresden,\\Helmholtzstr. 18, 01069 Dresden, Germany}

\author{Detlef Lohse}
\affiliation[UT]
{Physics of Fluids Group, Max Planck Center for Complex Fluid Dynamics,\\MESA+ Institute and J. M. Burgers Centre for Fluid Dynamics, University of Twente,\\P.O. Box 217, 7500AE Enschede, Netherlands}
\alsoaffiliation[MPI]
{Max Planck Institute for Dynamics and Self-Organization,\\Am Fassberg 17, 37077 G\"{o}ttingen, Germany}

\author{Dominik  Krug}
\affiliation[UT]
{Physics of Fluids Group, Max Planck Center for Complex Fluid Dynamics,\\MESA+ Institute and J. M. Burgers Centre for Fluid Dynamics, University of Twente,\\P.O. Box 217, 7500AE Enschede, Netherlands}

\author{Lars B{\"u}ttner}
\affiliation[TUD2]
{Laboratory for Measurement and Sensor System Techniques,\\Faculty of Electrical and Computer Engineering, Technische Universit{\"a}t Dresden,\\Helmholtzstr. 18, 01069 Dresden, Germany}

\author{Kerstin Eckert}
\affiliation[HZDR]
{Institute of Fluid Dynamics, Helmholtz-Zentrum Dresden-Rossendorf,\\Bautzner Landstrasse 400, 01328 Dresden, Germany}
\alsoaffiliation[TUD]
{Institute of Process Engineering and Environmental Technology,\\Technische Universit{\"a}t Dresden, 01062 Dresden, Germany}
\email{k.eckert@hzdr.de}

\title{Electrolyte spraying within H$_2$ bubbles during water electrolysis}

\begin{document}
\newpage
\begin{abstract}
Electrolytically generated gas bubbles can significantly hamper the overall electrolysis efficiency. Therefore it is crucial to understand their dynamics in order to optimise water electrolyzer systems. Here we demonstrate a distinct transport mechanism where coalescence with microbubbles drives electrolyte droplets, resulting from the fragmentation of the Worthington jet, into the gas phase during hydrogen evolution reaction, both in normal and microgravity environments. 
This indicates that the H$_2$ bubble is not only composed of hydrogen gas and vapor but also includes electrolyte fractions. Reminiscent of bursting bubbles on a liquid-gas interface, this behavior results in a flow inside the bubble, which is further affected by Marangoni convection at the gas-electrolyte interface, highlighting interface mobility. 
In the case of electrode-attached bubbles, the sprayed droplets form electrolyte puddles at the bubble-electrode contact area, affecting the dynamics near the three-phase contact line and favoring bubble detachment from the electrode. The results of this work unravel important insights into the physicochemical aspects of electrolytic gas bubbles, integral for optimizing gas-evolving electrochemical systems. Besides, our findings are essential for studying the limits of jet formation and rupture relevant to acid mist formation in electrowinning, generation of sea spray aerosols, impact of droplets on liquid surfaces, etc.
\end{abstract}

\newpage
The growth of gas bubbles abounds in nature and has various engineering applications\cite{lohse2018bubble} and is reflected in natural phenomena. Some of them featuring rapid dynamics are sonochemistry\cite{suslick1990sonochemistry} and sonoluminescence\cite{brenner2002single}, cavitation\cite{obreschkow2006cavitation}, the evolution of CO$_2$ bubbles in sparkling drinks \cite{liger2008recent}, and the bursting bubbles at the oceans surface \cite{villermaux2022bubbles, deike2022mass}. The latter contributes significantly to atmospheric aerosol generation\cite{blanco2020sea} via two different mechanisms: the disintegration of a thin liquid film between the bubble and gas interface at the onset of bursting \cite{lhuissier2012bursting,rage2020bubbles,jiang2024abyss}, and by an inertia-driven liquid jet--referred to as Worthington jet fragmenting into multiple droplets\cite{woodcock1953giant}, where the mechanism is a Rayleigh--Plateau instability\cite{ghabache2014physics}.
Beyond aerosol generation\cite{joung2015aerosol}, these jets are responsible for contaminant dispersion \cite{dubitsky2023enrichment,yang2023enhanced}. Additionally, they result in surface erosion and deformation through the impact of droplets on solid \cite{zhang2022impact} and liquid surfaces \cite{michon2017jet}, respectively.

A related problem also occurs in electrolysis, where the coalescence of hydrogen or oxygen bubbles can be approximated to bursting events at a liquid-gas interface. This is a particularly interesting problem of high practical relevance due to the prominent role of hydrogen produced via water electrolysis as an energy carrier, fuel, and feedstock for chemical and steel industries \cite{ staffell2019role}.
Alkaline water electrolysis is still the most mature technology, albeit suffering from inadequate efficiency when operated at high current densities. A considerable part of the losses can be attributed to the formation of H$_2$ and O$_2$ bubbles, present at the electrodes and in the bulk. These bubbles mask the active area of the electrodes, reduce the number of nucleation sites, and raise ohmic cell resistance \cite{angulo2020influence,hodges2022high}. Thus, enhanced removal of gas bubbles, inherently requiring a better understanding of their growth and departure, will promote continuous catalytic activity \cite{bashkatov2024performance} and benefit further optimization of the system's energy efficiency \cite{shih2022water}.

The dynamics of electrolytic bubbles have been extensively studied in the last few decades\cite{zhao2019gas,angulo2020influence} to uncover the growth laws controlled by either the interfacial diffusion of dissolved hydrogen\cite{raman2022potential,sepahi2024mass} or direct injection of the gas at the bubble foot\cite{bashkatov2022}; mass transfer and associated limitations \cite{haverkort2024general,sepahi2024mass}; interactions between neighboring bubbles\cite{bashkatov2024performance,demirkır2024}; the impact of the electrolyte composition\cite{park2023solutal}, also in the presence of surfactants\cite{fernandez2014bubble}; the force balance governing the bubble departure\cite{Hossain2022,demirkır2024} and finally the impact of bubbles on the cell overpotentials\cite{angulo2020influence, raman2022potential}. Only recently, the soluto- and thermocapillary Marangoni\cite{lubetkin2002motion,yang2018marangoni,massing2019thermocapillary,meulenbroek2021competing, park2023solutal} force and an electric\cite{bashkatov2019oscillating,Hossain2022} force caused by charge adsorption, which had not been considered before, have been uncovered and quantified. 
Furthermore, it has been discovered that H$_2$ bubbles on microelectrodes do not necessarily adhere to the surface. Instead, they might reside atop a "carpet" of microbubbles and grow via intensive coalescence with this bed of tiny precursors\cite{bashkatov2019oscillating,bashkatov2022}. However, the full implications of such rapid coalescence events in water electrolysis remain elusive--an area ripe for further inquiry. 
Several lingering questions are yet to be addressed: 
What are the main features of the coalescence in the confined geometry, set by H$_2$ bubble, carpet and electrode, and how do they interact with the Marangoni flow at the bubble surface? Under what conditions does electrolytic bubble coalescence lead to droplet and spray formation? Does this affect the contact line and potentially the detachment of the electrode-attached bubble?
In the present work, we address these open questions by combining experiments on the coalescence-driven dynamics of H$_2$ bubbles, focusing for the first time on the interior of the bubbles under both terrestrial and microgravity environments, alongside tailored direct numerical simulations.

\begin{figure}[h]
	\includegraphics[width=1\textwidth]{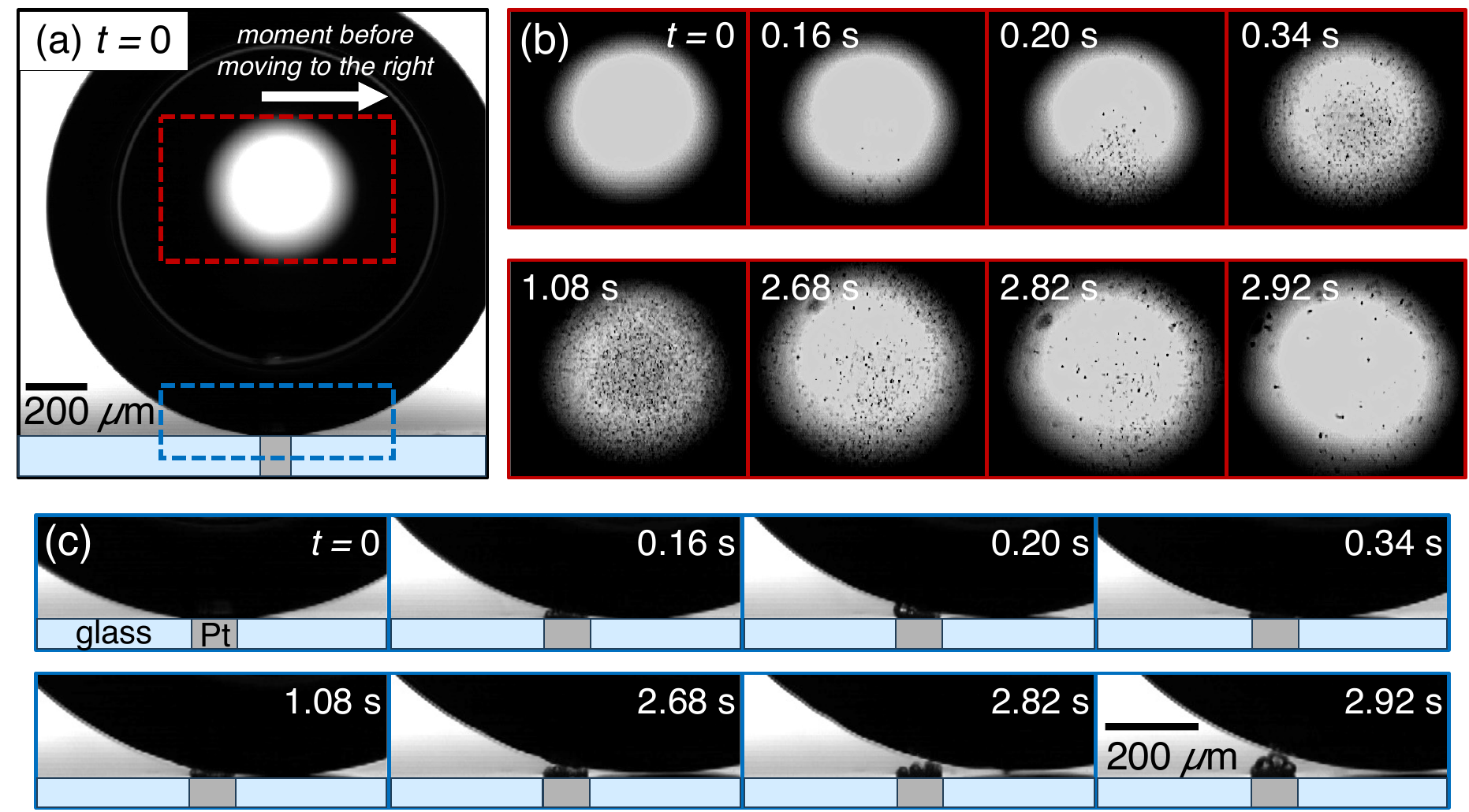}
	\caption{A series of shadowgraphs documenting the stream of electrolyte microdroplets inside a H$_2$ bubble ($R_b=902\:\mu$m at the departure) during the late phase of its evolution in a micro-$g$ environment. At $t=0$ in (a) the bubble sits at the electrode. The successive images in (b) and (c), zooming into central and lower segments of the bubble, demonstrate the emerging flow of electrolyte microdroplets, initiated soon after the onset of lateral motion to the right followed by the intensive coalescence events. The H$_2$ bubble is produced during water electrolysis at $100\:\mu$m Pt micro-electrode at $-4$~V (vs Pt wire) in 0.5 mol/L H$_2$SO$_4$. The image recording was performed with a frame rate of 50 Hz. 
	\label{fig:fig1}}
\end{figure}

The main phenomenon under study, 
spray formation {\it inside} a H$_2$ bubble during water electrolysis, is presented in Figure \ref{fig:fig1}. This observation was made under microgravity conditions provided by parabolic flights of an Airbus A300\cite{bashkatov2021dynamics}.
The snapshot at $t=0$ shown in Fig.~\ref{fig:fig1}a documents the time instant when the bubble sits at the electrode surface, blocking most of its active area, hindering the reaction and hydrogen production rates.
Figures~\ref{fig:fig1}b,c respectively zoom into the bubble's central and lower segments over various time points leading up to its departure.
Soon after $t=0$, the bubble begins a lateral shift to the right driven by residual gravitational forces 
releasing the electrode and enabling the formation of a dense carpet of microbubbles. As a result, the primary bubble continuously coalescences with these microbubbles emerging on a time scale of $\mathcal{O}$($\mu$s).
The successive images document an emerging flow consisting of electrolyte droplets, which is initiated soon after the onset of coalescence events and ascends from the base of the bubble
toward its apex. These droplets become noticeable at $t = 0.16$~s, with their population density peaking at $t = 1.08$~s and declining by $t=2.92$~s. The gradual widening of the gap between electrode and bubble interface over time has two different effects:
(i) It enhances the electrochemical reaction (by increasing the electric current, see Supplementary Section 1), thereby elevating H$_2$ production and bubble--carpet coalescence rates, and (ii) it leads to the generation of larger pre-coalescence bubbles, in turn decreasing the frequency of the coalescence events and, subsequently, droplet injections into the main bubble. The competition between these two effects establishes an optimal carpet thickness at which the coalescence rate has its maximum. Beyond this distance, the droplet population significantly reduces, as evidenced at  $t = 2.92$~s. 
The droplet radii remain approximately constant at 1.8 $\pm$ 0.8~$\mu$m during most of the coalescence phase, increasing to about 3.1 $\pm$ 1.3~$\mu$m only just before bubble departure, when the gap between bubble interface and electrode is at its maximum.

\begin{figure}[h]
	\includegraphics[width=1\textwidth]{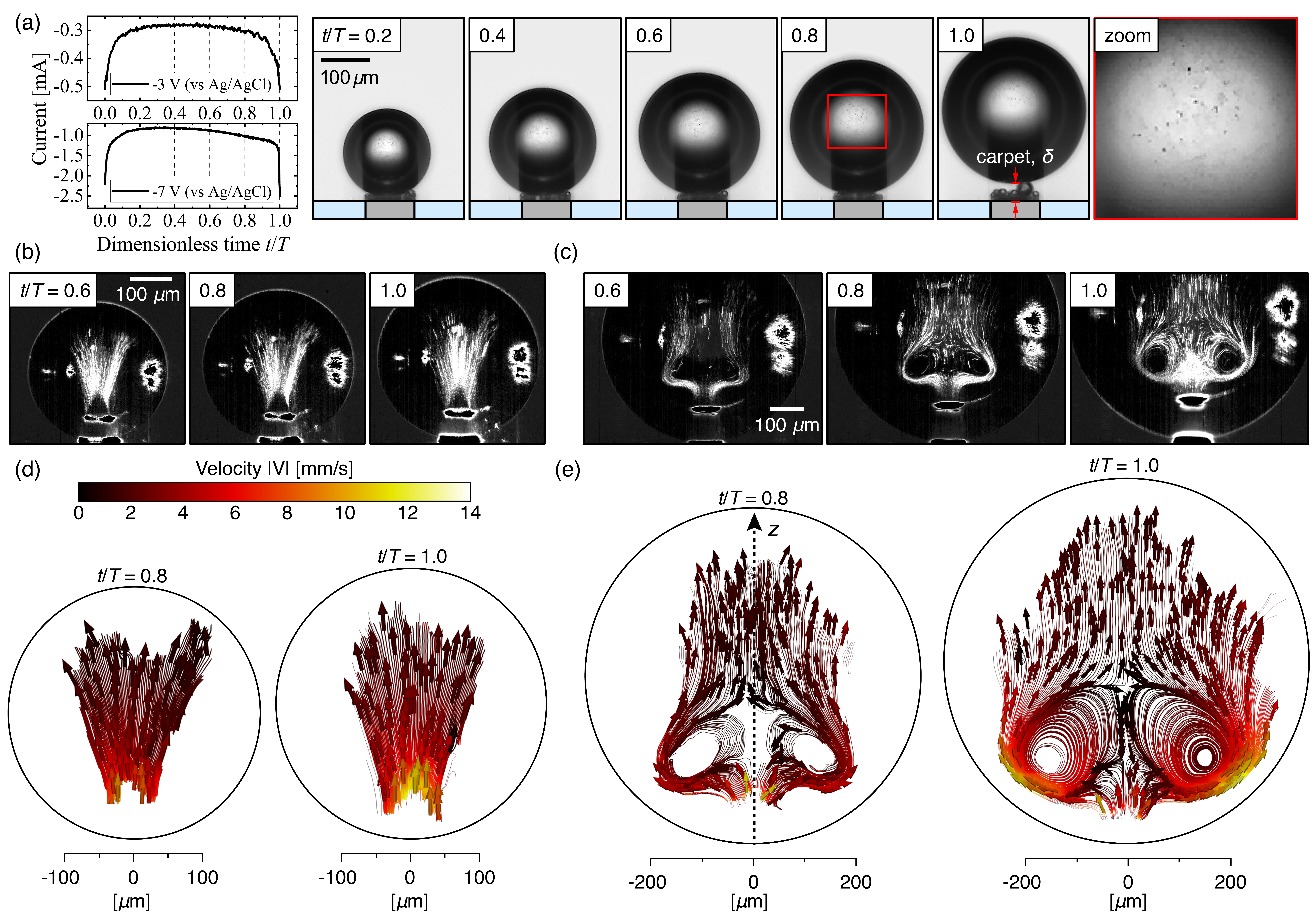}
	\caption{The dynamics of H$_2$ bubble presented in terms of (a) electric current at $-3$~V (top) and $-7$~V (bottom), supplemented with shadowgraphs throughout its evolution at $-3$~V. The most right image zooms into the middle part of the bubble at $t/T=0.8$. (b-c) Snapshots highlighting the streaklines of the droplets over $\Delta$$t$. (d-e) The streamlines of the averaged drop velocity field. 
 For the velocity calculations, the optical distortions (aberration) caused by the curvature of the bubble are corrected analytically, before the velocity calculations (Supplementary Section 2). The measurements in (b-d) and (c-e) performed in 0.1 mol/L H$_2$SO$_4$ at $-3$~V and $-7$~V vs Ag/AgCl, respectively.
		\label{fig:fig2}}
\end{figure}

In the following, we demonstrate how the phenomenon manifests itself under normal gravity conditions. 
Figure \ref{fig:fig2} illustrates (a) the electric current $I$ at $-3$ V and $-7$ V, and shadowgraphs along bubble evolution at $-3$~V in 0.1~mol/L H$_2$SO$_4$.
In detail, a single primary bubble forms via coalescence shortly after nucleation at $t/T=0$ from many nano- and micrometer bubbles\cite{bashkatov2022}. It continues to grow through rapid $\mathcal{O}$($\mu$s) coalescence with the carpet of microbubbles beneath.
The evolution ends with the bubble departure at $t/T = 1$ when buoyancy overcomes downward forces \cite{bashkatov2022,Hossain2022}. $T$ is the bubble lifetime.
$I(t)$ reflects variations in ohmic resistance due to bubble size and position relative to the electrode, peaking between the departure and nucleation of the next bubble.

In analogy to Fig. \ref{fig:fig1}, numerous electrolyte droplets are injected during the coalescence events, as seen in the last image of Fig. \ref{fig:fig2}a, which focuses on the central segment of the bubble at $t/T =0.8$.
The snapshots in (b,c) highlight the streaklines of the droplets over $\Delta t = 25$ ms, emerging at the bubble-carpet interface and moving towards the bubble apex with the velocities plotted in (d,e).

The flow in Fig.~\ref{fig:fig2}b-e develops continuously throughout the bubble evolution, along with and in response to the growing carpet thickness\cite{bashkatov2022,park2023solutal} and hence elevated current, reaching velocities of up to 14 mm/s at $t/T = 1.0$.
High-speed recordings at 600 kHz and 720 kHz (Supplementary Section 3) reveal that some droplets are injected at velocities up to 15.8 m/s, i.e. three orders of magnitude higher. These rare events, resulting in larger droplets, occur around the bubble's departure when the carpet thickness is at its maximum, approximately between $\delta=16\:\mu$m and $\delta=43\:\mu$m (see Fig. \ref{fig:fig2}a), but are not observed during the earlier stages of the bubble's evolution when $\delta<16\:\mu$m.

At a substantially larger electric current (see Fig. \ref{fig:fig2}a),  the flow is altered by the presence of a vortical structure, see a transition from a fireworks-like shape (b,d) at $-3$~V to a vortex-like shape at $-7$~V (c,e). Meanwhile, the flow at the base of the injection remains similar. At lower potential, the flow expands away from the injection source, while at higher potential, the droplets are carried away from the injection area and ascend along the bubble-electrolyte interface. In the latter, some droplets enclosing the vortex are carried back toward the electrode. In both cases, the velocity gradually decays with distance. Indeed, the velocity of injected droplets is expected to decay exponentially over time due to viscous drag: 
$V_{d}(t) = V_0 \cdot \exp\left(-t/\tau_d
\right)$
(see Supplementary Section 4). Here, $\tau_d = \frac{m_d}{6 \cdot \pi \cdot \mu_{H2}\cdot R_{d}}$ and $R_d$ is the radius of the droplet, $m_d$ is the mass of the droplet,
$V_0$ is the initial velocity and $\mu_{H2}$ is the dynamic viscosity of H$_2$. For example, a droplet with $R_d$ = 1~$\mu$m (implies $\tau_d \approx 25$~$\mu$s) and initial velocity $V_0 = 5$~m/s slows to $10^{-2}$~m/s in just 160~$\mu$s by traveling 124~$\mu$m. At $-7$~V, droplets near the \textit{z}-symmetry line (see Fig. \ref{fig:fig2}e) are dragged into the downward flow stream, enclosing the vortex.

The flow transition observed between $-3$~V and $-7$~V is due to Marangoni convection around an electrogenerated gas bubble existing at its outer interface. 
This convection originates from a gradient of surface tension caused by thermo- and/or solutocapillary effects \cite{massing2019thermocapillary,yang2018marangoni,park2023solutal}, creating a shear stress imbalance that moves the fluid-gas interface.
The resulting flow is directed alongside the electrolyte-gas interface from small to large values of surface tension, i.e., from the bottom to the top of the bubble. These effects are localized at the foot of the bubble and are consistent with the position of the vortex ring in Fig.~\ref{fig:fig2}e.
Thermal Marangoni forces are driven by Joule heating from locally high current density ($j$) at the wetted part of the electrode (Fig. \ref{fig:fig5}b) and scale (via Ohm's law) with $j^2$, while solutal Marangoni forces arise from electrolyte depletion at the electrode and depend linearly on $j$. At higher potentials, as in the present study, the Marangoni convection is mainly driven by thermal effects \cite{park2023solutal}, with temperature rising up to 14~K \cite{massing2019thermocapillary}. The velocity magnitude scales with the electric current \cite{yang2018marangoni} and may reach about 10~mm/s at $-2.2$~mA and 47~mm/s at $-4.8$~mA in 0.5~mol/L H$_2$SO$_4$ \cite{babich2023situ}. This concludes that the pronounced variance in flow structure between $-3$~V and $-7$~V originates from the substantial difference in electric current magnitude, and consequently, the Marangoni convection.
Thus, reminiscent of evaporating droplets\cite{lohse2020physicochemical} or rising bubbles\cite{young1959motion}, Fig.~\ref{fig:fig2} demonstrates for the first time that Marangoni convection at the electrolyte-gas interface drives internal flow in electrogenerated gas bubbles, directing and accelerating injected microdroplets. This also indicates that the gas-electrolyte interface is mobile, though the mechanism behind preferential ion adsorption and its effects remain unclear.

\begin{figure}[h]
	\includegraphics[width=0.9\textwidth]{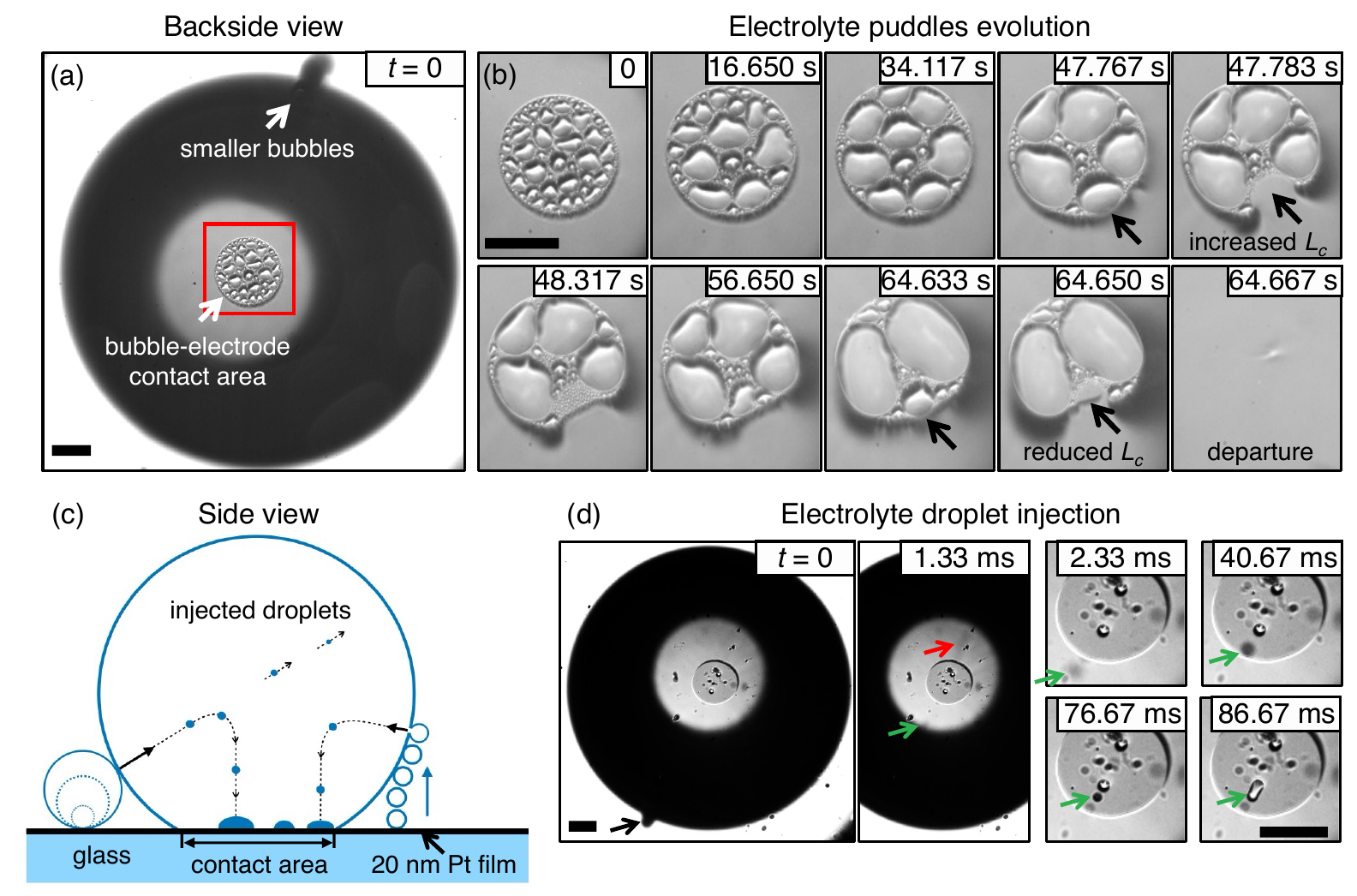}
	\caption{(a,b,d) Backside views from underneath the electrode of the growing H$_2$ bubble attached to the transparent planar Pt electrode. (b) Zooms into the bubble-electrode contact area shown by the red rectangular in (a), demonstrating the development of electrolyte puddles throughout the bubble evolution. (c,d) Schematic and shadowgraphs illustrating the injection of microdroplets upon coalescence events followed by their sedimentation at the contact area. Scale bars are 100 $\mu$m. The measurements were carried out at a current density of 50 $\mathrm{A/m^2}$ in 0.1 mol/L HClO$_4$. The image recording in (a,b) and (d) was performed at frame rates of 60 Hz and 3000 Hz, respectively. 
		\label{fig:fig3}}
\end{figure}

Another intriguing outcome of the spraying, shown in Figure \ref{fig:fig3}, is the formation of electrolyte fractions within an electrode-attached and growing H$_2$ bubble, specifically at the contact area with the electrode surface\cite{fernandez2014bubble,demirkır2024}.
Figure~\ref{fig:fig3}a,b,d documents the views from below a transparent planar electrode (20 nm of Pt).
The snapshots in Fig. \ref{fig:fig3}b zoom in on the contact patch (area marked by the red rectangular in Fig. \ref{fig:fig3}a), which is seen to feature sessile electrolyte droplets inside the gas phase, that 
expands throughout the bubble evolution\cite{demirkır2024}. The bubble grows mainly due to diffusion of the dissolved gas but also via coalescence with the neighboring bubbles. Here, the smaller bubbles nucleate below the equator of the primary bubble and quickly detach, see a plume of out-of-focus small bubbles in Fig. \ref{fig:fig3}a, likely due to the altered morphology/wettability of a tiny cavity they nucleated at. Consequently, upon reaching the gas-liquid interface of the larger bubble, coalescence occurs between the two, see schematic in Fig.~\ref{fig:fig3}c.

Figure \ref{fig:fig3}d details the injection of at least two microdroplets, marked by red and green arrows, following the coalescence event between the primary and smaller bubble (black arrow at $t=0$). 
The first droplet (red arrow) moves with a much faster velocity, likely shooting through the gas-electrolyte interface on the opposite side. In contrast, the second droplet, about $r=4$ $\mu$m (green arrow), slows down quickly due to Stokes's drag (Supplementary Section 4) and falls, presumably at terminal velocity, to the contact patch at $t = 86.67$~ms, merging with another droplet. -- In detail, it moves with an average velocity of about $\overline{V}_d = 0.26$ m/s within the first 1.33 ms and about $\overline{V}_d = 6$ mm/s between 1.33 ms and 86.67 ms, assuming the traveled distance $S_d$ equals the bubble radius $R_b=509$ $\mu$m. The latter correlates well with the terminal velocity of the droplet $V_t = 4.1$ mm/s in the Stokes regime (Supplementary Section 4).
The process repeats during numerous coalescence events, resulting in the gradual formation of electrolyte puddles as shown in Fig.~\ref{fig:fig3}b. These puddles grow in size throughout the bubble evolution, as more electrolyte droplets are injected, wetting larger areas of the electrode. 
Once any of the puddles reaches the gas-electrolyte interface, it rapidly merges with the electrolyte bulk, thereby moving the contact line and effectively reducing the bubble-electrode contact area (see frames at 47.783~s and 64.650~s). 

This process thus plays a key role for the bubble detachment. The detachment size of an electrode-attached bubble is primarily governed by the surface tension force $F_s$, which depends on the length of the contact line ($L_c$). Comparing the snapshots at 47.783 s and 64.650 s, the length of the contact line can either increase or decrease after the puddle merges into the electrolyte bulk. A sudden reduction in $L_c$, provided there is sufficient buoyancy, causes an earlier detachment from the electrode surface, as illustrated in the snapshot at 64.667~s. The scarcity of electrolyte puddles in (d) is attributed to the reduced number of nucleation sites and their lower activity near the primary bubble, resulting in a lower frequency of coalescence events and fewer injected droplets.

\begin{figure}[h!]
\includegraphics[width=0.9\textwidth]{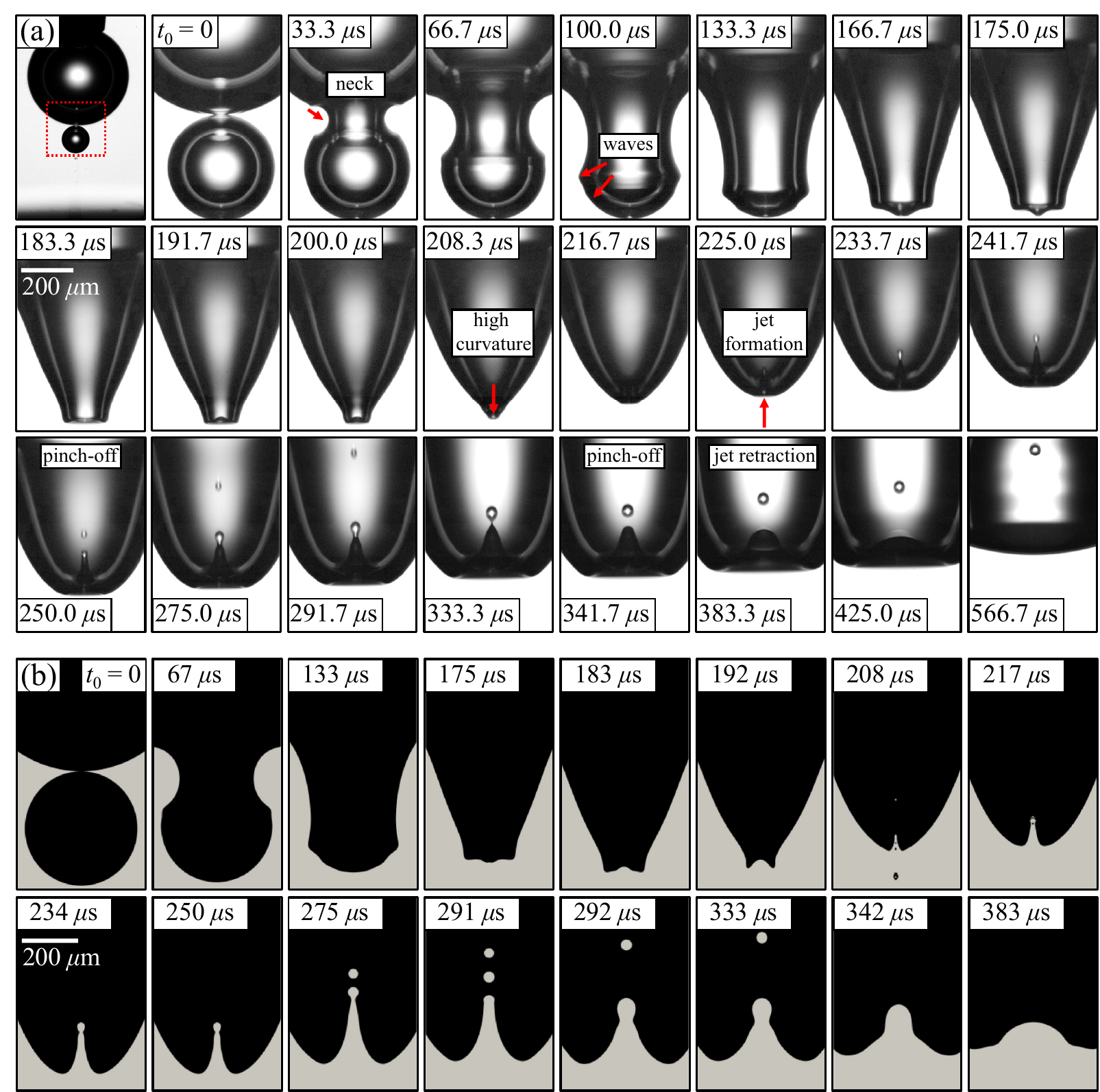}
\caption{Droplet ejection mechanism upon coalescence of two unequal size H$_2$ bubbles shown by snapshots from (a) experiment and (b) numerical simulation. In the experiment, both bubbles were produced during electrolysis in 0.5 mol/L H$_2$SO$_4$. While the bigger bubble is pinned to a blunt needle, the smaller bubble rises from the electrode until the coalescence begins at $t_0=0$. The coalescence process is accompanied by the injection of two droplets after their consecutive separation from the jet. The first snapshot in (a) demonstrates the configuration marking the region of interest by the red square. The image recording was performed at 120 kHz.
\label{fig:fig4}}
\end{figure}

Figure \ref{fig:fig4}a shows a sequence of shadowgraphs detailing the mechanism of droplet injection characterized by the formation of an internal jet that entrains a volume of electrolyte, known in the fluid mechanics and physical oceanography communities as the Worthington jet \cite{worthington1877}. The process is demonstrated by two coalescing H$_2$ bubbles with sizes $R_b = 400\:\mu$m and $R_s = 205\:\mu$m, respectively. The results are corroborated by direct numerical simulations (DNS) shown in Fig. \ref{fig:fig4}b. 
In detail, when a smaller bubble touches a larger one, the liquid film that separates the bubbles gradually drains, forming a neck connecting the two ($t=33.3$~$\mu$s)\cite{eggers2024coalescence}. Growth of this neck follows a Taylor--Culick-type mechanism\cite{taylor1959dynamics,culick1960comments,munro2015thin} exciting capillary waves that propagate along the bubble interface \cite{deike2018dynamics,gordillo2019capillary}, see $t=66.7\:\mu$s to $183.3$~$\mu$s. 
The viscous forces dictate the motion of these capillary waves enervating all but the strongest (with highest curvature) waves that ultimately focus at the bottom and induce a region of high curvature \cite{duchemin2002jet,walls2015jet,sanjay2021bursting}, see $t=191.7\:\mu$s to 208.3~$\mu$s. This inertial flow focusing creates an upward jet ($t=216.7\:\mu$s to 241.7~$\mu$s)\cite{gordillo2019capillary,sanjay2021bursting,gordillo2023theory} propagating inside of the merging H$_2$ bubbles. Consequently, this process is controlled by the dimensionless viscosity of the electrolyte given by the Ohnesorge number $Oh$
\begin{equation}
    \label{eq:Oh}
    Oh = \frac{\mu_{el}}{\sqrt{\rho_{el}\gamma R_s}},
\end{equation}
\noindent where $\mu_{el}$ represents the dynamic viscosity, $\rho_{el}$ the density of the electrolyte, $\gamma$ the surface tension, and $R_s$ the initial radius of the smaller bubble.
Eventually, the jet breaks into two droplets due to the Rayleigh--Plateau instability\cite{sanjay2021bursting,walls2015jet}, for $Oh < Oh^*$ where $Oh^* \approx 0.035$ is the critical Ohnesorge number for drops--no-drops transition\cite{sanjay2021bursting,walls2015jet} for bursting at the liquid-gas free interface.  
Beyond the critical Ohnesorge number, viscous dissipation dominates,
ceasing the ejection of drops. On further increasing the Ohnesorge number ($Oh > 0.1$), the Worthington jet does not form\cite{walls2015jet, sanjay2021bursting}.

The DNS results in Fig. \ref{fig:fig4}b accurately reproduce key features and timescales of the phenomenon such as neck formation, capillary wave propagation, formation, and breakup of the jet. 
In the experiments, the first droplet with a radius of \( r_d = 13 \, \mu \text{m} \) is observed at \( t = 250.0 \, \mu \text{s} \) and ejects with a velocity of approximately \(\overline{V}_d = 7.2 \, \text{m/s} \). In close qualitative and quantitative agreement, the simulation demonstrates the first droplet (radius \( r_d = 15 \, \mu \text{m} \)) pinching off at \( t = 260 \, \mu \text{s} \) with \( \overline{V}_d = 4.3 \, \text{m/s} \).
Further details and discussion on the origins of minor quantitative discrepancies are elaborated in Supplementary Section 5.

It is important to note that the injection demonstrated at \(Oh = 0.008\) in Fig. \ref{fig:fig4} represents a relatively isolated but conventional case \cite{walls2015jet,sanjay2021bursting}, with the smaller bubble being located far from the electrode.
In contrast, the bursting events in Figs. \ref{fig:fig1} and \ref{fig:fig2} taking place in a highly confined configuration near the Pt electrode feature high coalescence rates and involve smaller bubbles (up to about \( R_s = \delta/2 = 8 \, \mu \text{m} \)). Despite the higher Ohnesorge number ($Oh$ = 0.042), injections still occur, exceeding the critical $Oh^*$ found for an unconfined isolated bubble.
This observation suggests that a nearby wall and high coalescence rates can significantly influence the injection mechanism.
In agreement with this, Lee et al. (2011)\cite{lee2011size} identified a higher critical value $Oh^* = 0.052$, specifically for smaller bubbles ($Bo < 10^{-3}$) bursting near a solid boundary. We refer the readers to Sanjay et al. (2024)\cite{sanjay2024asymmetry} for further details. 
	
Lee et al. (2011)\cite{lee2011size} also studied a bubble with a relatively small $R_s$ = 26.5~$\mu$m adjacent to a Pt substrate using ultrafast X-ray imaging, finding daughter aerosol droplets (2~$\mu$m to 4~$\mu$m radii) with velocities around 0.3 m/s (Supplementary Movie 5 in \citet{lee2011size}). 
Consequently, we can classify the bursting events in order of increasing droplet speed: (i) carpet bubbles ($R_s=8$ $\mu$m) bursting near a solid wall with a velocity of $V_d\sim 10^{-2}$ m/s, (ii) a bubble with $R_s=26.5$ $\mu$m bursting near the solid wall, resulting in a droplet speed of $V_d\sim 10^{-1}$ m/s, and (iii) the bubble with $R_s=205$ $\mu$m bursting away from the wall (as detailed in Fig. \ref{fig:fig4}), which results in a droplet speed in the range of $V_d\sim 10^0$ m/s to $10^{1}$ m/s. 
Further deceleration likely comes from viscous drag within the surrounding H$_2$ gas, as described by the Oseen approximation to the Stokes flow\cite{book-lamb}. 
Finally, a high coalescence rate, as seen in Figs. \ref{fig:fig1} and \ref{fig:fig2}, could disrupt the symmetry of coalescence, affecting the propagation of capillary waves in each event and potentially significantly reducing the velocity of the ejected droplets to $V_d\sim 10^{-2}$ m/s. 
Therefore, the small initial size of the bursting bubble (i.e., large $Oh$), proximity to a wall, higher viscosity of the gas bubbles, and potentially high coalescence rates can substantially reduce the injection velocity.

Our findings demonstrate a distinct transport mechanism of electrolyte droplets inside the gas phase during water electrolysis.
As discussed above, the coalescence of a primary bubble with the bubbles-satellites causes the electrolyte spraying via the fragmentation of the Worthington jet. This indicates that the H$_2$ bubble is not only composed of hydrogen gas and vapor but includes electrolyte fractions given the coalescence with nearby bubbles.
We emphasize again that the microdroplets formed in the bubble through this process play an important role for the bubble detachment, once they merge with the surrounding electrolyte at the contact line.
The results we report will be integral for further studying the limits of jet formation and rupture associated with $Oh^*$ in confined geometries near a solid boundary. 
Additionally, our findings will be valuable for validating and tailoring numerical and theoretical models.
We highlight that the injected droplets serve as a non-invasive tool, making the internal flows associated with Marangoni convection at the electrolyte-gas interface visible and quantifiable for the first time. 
This gives access to the important surface mobility of electrogenerated bubbles, which is determined by preferential ion adsorption --- a phenomenon that remains poorly understood.
This will allow to access the role of physicochemistry in the hydrodynamic phenomena related to bubbles. The knowledge could further be transferred to the other side of the electrochemical reaction --- the formation of O$_2$ bubbles.
The results of this work unravel important insights into the physicochemical aspects of electrochemically generated H$_2$ gas bubbles and have broad relevance, e.g. to acid mist formation in electrowinning processes\cite{ma2020characteristics}; the generation of sea spray aerosols\cite{villermaux2022bubbles}, which play a role in airborne disease and pollutant transmission\cite{yang2023enhanced}; bursting CO$_2$ bubbles in sparkling drinks\cite{liger2008recent}; and to the impact of droplets on liquid\cite{michon2017jet} surfaces. 
In particular, the findings are essential for the water electrolysis field, where a deeper understanding of bubble evolution mechanisms is essential for optimizing gas-evolving electrochemical systems.

\newpage
\section{Methods}
\label{sec:methods}
The hydrogen gas bubbles were produced using both micro- and planar electodes during water electrolysis.
Part of the results (see Fig. \ref{fig:fig1}) were obtained in a microgravity environment achieved during the 34th DLR Parabolic Flight Campaign in September 2019 (see Bashkatov et al.\cite{bashkatov2021dynamics}).

\subsubsection{Microelectrode system}
\begin{figure}
	\includegraphics[width=1\textwidth]{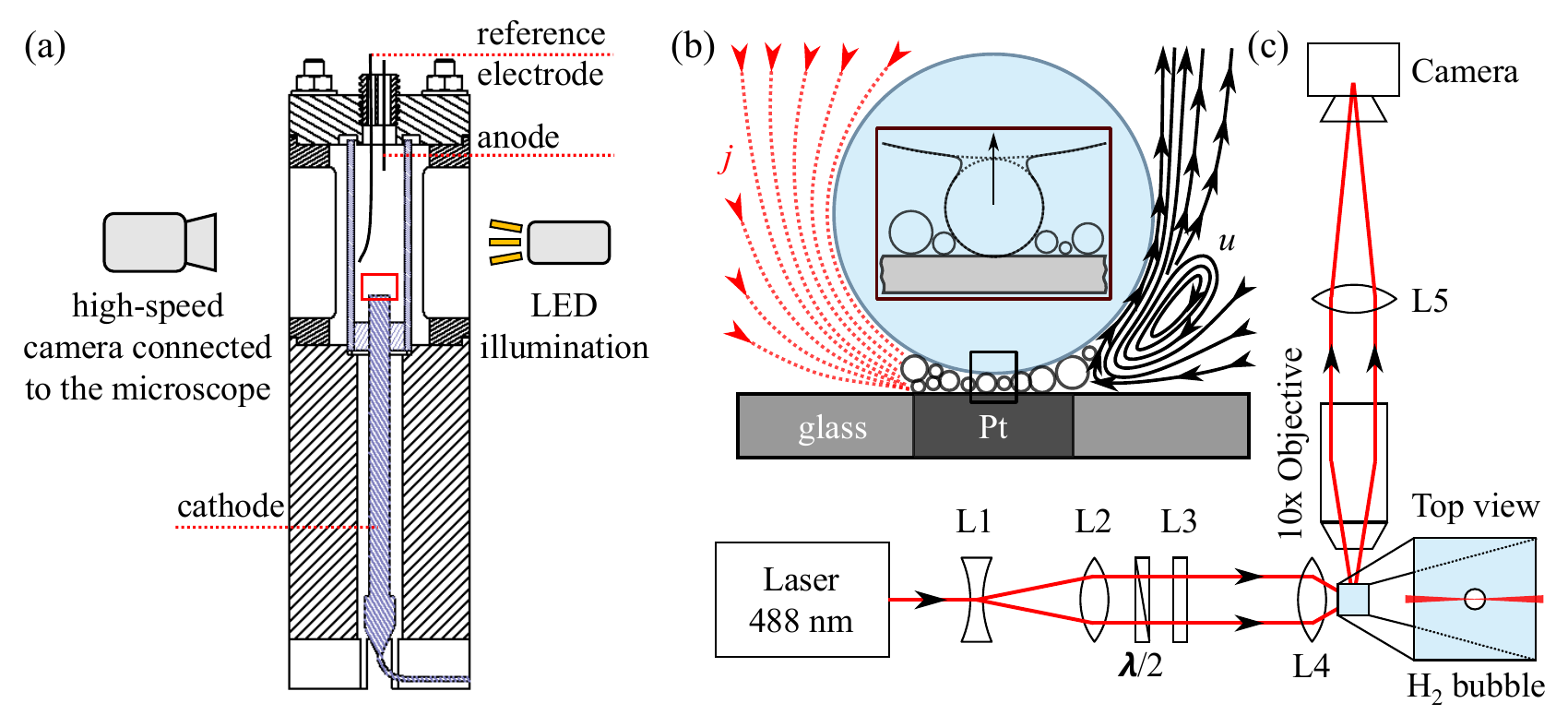}
	\caption{Schematics (not to scale) of (a) an electrochemical cell and a shadowgraphy system; (b) H$_2$ bubble sitting on the carpet of microbubbles generating between its bottom and electrode surface. Inset zooms into the bottom of the bubble where an intensive bubble-carpet coalescence takes place; (c) PTV optics used to measure the velocity of the injected droplets inside the H$_2$ bubble. For details see text.
		\label{fig:fig5}}
\end{figure}

Single hydrogen gas bubbles growing on the carpet of microbubbles were produced using a three-electrode electrochemical cell filled with sulfuric acid of either 0.1~mol/L or 0.5~mol/L concentration, see Fig.~\ref{fig:fig5}a. It comprises a cathode (Pt microelectrode, $\diameter$100 $\mu$m, ALS Co., Ltd) inserted horizontally facing upward in the base of a transparent cuboid glass cuvette (Hellma) with dimensions of 10 $\times$ 10 $\times$ 40~mm$^3$, anode (Pt wire, $\diameter$0.5~mm) and a reference electrode (Ag/AgCl) both inserted from the top. The experiments in a microgravity environment were done using a pseudo-reference electrode (identical to the anode)\cite{bashkatov2021dynamics}. The electrochemical cell was fixed inside an outer housing featuring two optically accessible observation windows. Before the measurements, the microelectrode surface underwent mechanical polishing with sandpaper (2000 grit), sonication, and rinsing with ultrapure water. For microgravity experiments, it was polished by diamond (1 $\mu$m) and alumina (0.05~$\mu$m) suspensions (ALS Co., Ltd) instead. The cell was connected to an electrochemical workstation (CHI 660E) and operated at a constant potential of either $-3$~V, $-4$~V or $-7$~V.

The experiments using a blunt needle in Figure \ref{fig:fig4} were performed as follows. First, a larger H$_2$ bubble with a radius of approximately 400~$\mu$m was generated at the microelectrode and then detached following a potential interruption. As it rose, it adhered to a blunt needle positioned above the microelectrode, with surface tension keeping the bubble attached. A second, smaller bubble with a radius of approximately 205 $\mu$m was created in the same manner, with the smaller size achieved by applying a shorter pulse of potential. As this smaller bubble rose, it contacted the larger bubble, initiating the coalescence process. The time $t_0$ marks the moment just before coalescence begins.

\subsubsection{Planar electrode system}
The electrode-attached hydrogen gas bubbles were produced at the surface of a $\diameter$50 mm disc-like planar electrode (cathode) inserted horizontally facing upward in the base of the cylindrical PTFE compartment with an inner diameter of 40 mm and a height of 50 mm filled with 0.1 mol/L HClO$_4$. The cathode was fabricated by sputtering a 20 nm thin film of platinum onto a glass slide, with a 3 nm tantalum layer applied for improved adhesion. The thin layer of platinum ensured the transparency of the cathode and allowed the visualization from the bottom of the cell. The cell was completed by a platinized titanium mesh (anode) and the Ag/AgCl reference electrode both inserted from the top. The system was controlled by the electrochemical workstation (Biologic VSP-300) maintaining a constant current density of 50 $\mathrm{A/m^2}$. The relatively low current density and smooth surface of the cathode allowed only a limited number of active nucleation sites, making the study of the contact line and electrolyte puddles dynamics possible. For details, we refer to \citet{demirkır2024}.

\subsubsection{Shadowgraphy system}
The visualization of the bubble dynamics is performed using a conventional shadowgraphy system, shown schematically for a microelectrode system in Figure \ref{fig:fig5}a. It consists of a high-speed camera connected to the microscope and LED illumination. 
The shadowgraphs in Figs. \ref{fig:fig1} and \ref{fig:fig2}a were recorded using an IDT camera (NX4-S1 and Os7-S3) with spatial resolutions of 678 pixels/mm and 1000 pixels/mm, respectively. In Figs. \ref{fig:fig3} and \ref{fig:fig4}a, a Photron camera (FASTCAM NOVA S16) was used, with spatial resolutions of 530 pixels/mm and 496 pixels/mm, respectively.
To achieve the bottom view (planar electrode system), the optical path of a horizontally installed camera is redirected vertically through the transparent cathode using a 45$^{\circ}$ mirror mounted below the electrode\cite{demirkır2024}. The LED light illuminates perpendicularly to the electrode from the top of the cell. The vertical adjustments of the focal plane are achieved using a high-precision motorized stage.

\subsubsection{Particle Tracking Velocimetry (PTV)}
The evolution of H$_2$ bubbles at microelectrodes is featured by the intensive coalescence with the carpet of microbubbles sandwiched between the bubble bottom and electrode (Fig. \ref{fig:fig5}b) on a time scale of $\mu$s. Owing to these coalescence events, multiple electrolyte droplets are injected into the bubble. The velocity measurement of these electrolyte droplets is performed using a Particle Tracking Velocimetry (PTV) system, schematically shown in Figure \ref{fig:fig5}c.

The setup employs a light sheet optical configuration comprising a laser (OBIS 488LX, 150~mW, Coherent Inc.) that was spatially enlarged using a telescope (L1 \& L2). To minimize reflection at the bubble surface, a $\lambda/2$-waveplate is employed to rotate the polarization. Subsequently, the beam is vertically expanded using a cylindrical lens (L3) before being focused inside the bubble by another lens (L4) with a focal length $f=19$~mm. For imaging purposes, a microscope objective (PLN 10X, Olympus) is positioned such that the bubble resides within the working distance. Finally, the bubble is imaged onto the camera (EoSens 3CXP, Mikrotron) using a lens (L5) with a focal length $f=160$ mm. To resolve the contours of the bubble, the system additionally possesses a background LED illumination. A series of images from Fig.~\ref{fig:fig2}b-c is collected at 1 kHz having a spatial resolution of 1140~pix/mm. The resulting series of images were processed by the software DaViS 10, which employs a Particle Tracking Velocimetry (PTV) algorithm to track each particle (droplet) over 25~ms at $t/T = 0.6$, 0.8 and 1.0. Due to the limited number of droplets, the resulting tracks were collected over several bubbles. Subsequently, the tracks were converted into a vector field using a binning function that interpolates local tracks on a specified fine grid. Finally, the vector fields are used to plot the streamlines of the averaged drop velocity field shown in Fig. \ref{fig:fig2}.

\subsubsection{Numerical method}
\begin{figure}
	\includegraphics[width=0.45\textwidth]{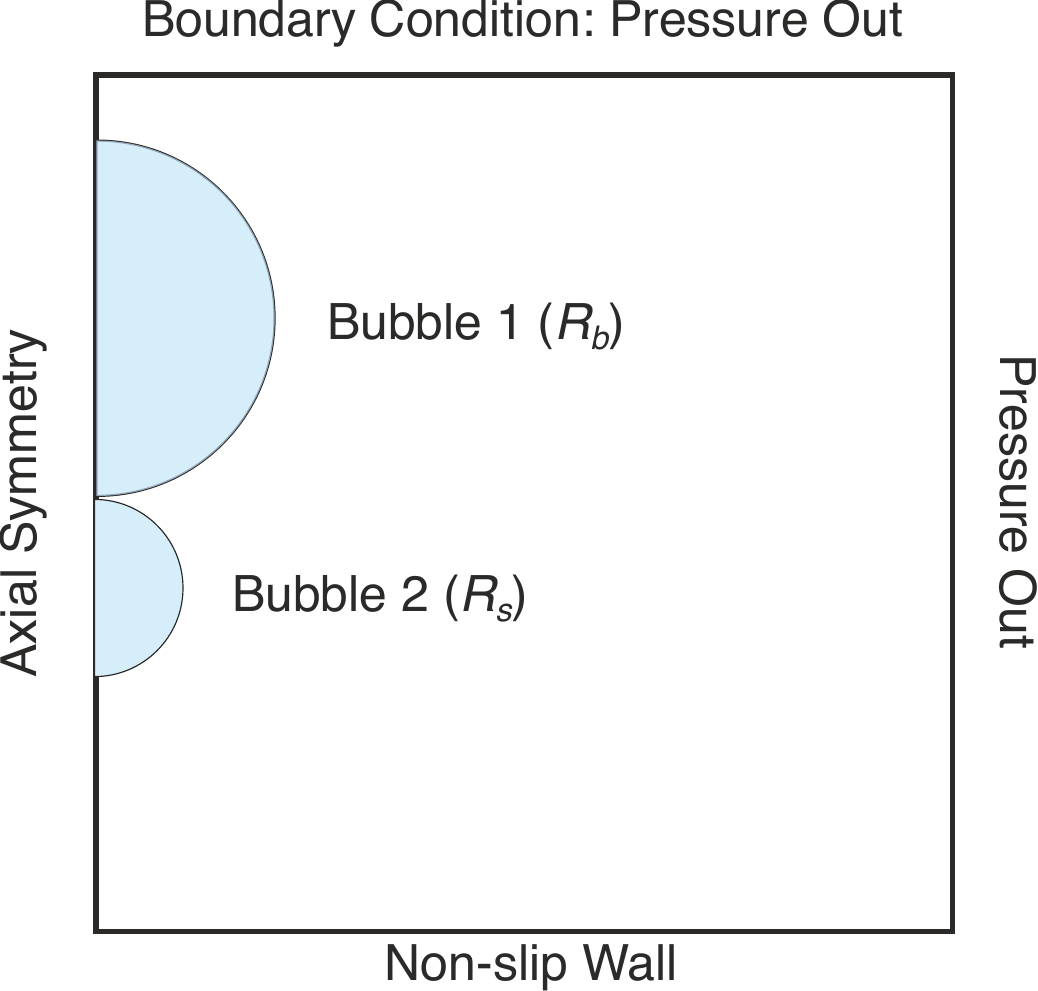}
	\caption{Sketch of the simulation model.
		\label{met:sim}}
\end{figure}

In this work, the direct numerical simulation code, Basilisk, is employed to simulate the coalescence of two bubbles. A two-fluid model, combined with a Navier Stokes solver, is employed. The interface of the liquid and gas is tracked with the Volume of Fluid (VOF) method. The liquid phase is water with a density and dynamic viscosity of 1000~kg/m$^3$ and 0.00105~Pa$\cdot$s, respectively. The gas phase is air, with a density and dynamic viscosity of 1.41~kg$\cdot$m$^3$ and 1.46 $\cdot$ $10^{-5}$ Pa$\cdot$s. The surface tension on the interface of liquid and gas is 0.072~N/m. The initial radius of bubble 1 is $R_b = 400\:\mu$m and bubble 2 is $R_s = 200\:\mu$m. Figure~\ref{met:sim} demonstrates a sketch of the simulation model.

Spatial discretization is performed using a quad-tree method in a 2D axisymmetric calculation domain of 1.5$\cdot$10$^{-3}$~m $\times$ 1.5$\cdot$10$^{-3}$~m. The adaptative Mesh Refinement algorithm was used to increase the calculation accuracy and reduce the hardware requirement. The maximum refinement level and the minimum level are 9 and 5, respectively. The calculation time step size is set to 1$\cdot$10$^{-8}$ s. 

\section{Acknowledgements}
This research received funding from the German Space Agency (DLR), with funds provided by the Federal Ministry of Economics and Technology (BMWi) due to an enactment of the German Bundestag under Grant No. DLR 50WM2352 (project MADAGAS III), H2Giga (BMBF, 03HY123E), from the Hydrogen Lab of the School of Engineering of TU Dresden, from the Advanced Research Center Chemical Building Blocks Consortium (ARC CBBC), under the project of New Chemistry for a Sustainable Future (project number 2021.038.C.UT.14) and partially from the German Research Foundation (DFG, project number 459505672).

\section{Author contributions}
A.B. and K.E. conceived the project. 
A.B., F.B., Ç.D., A.B., X.Y., D.L., D.K., L.B. and K.E. designed the experiments. 
A.B., F.B., Ç.D. and X.Y. carried out the experiments.
W.D. and V.S. carried out numerical simulations. 
A.B., Ç.D., W.D., V.S., A.B., G.M., J.C, D.L., D.K. and K.E. carried out bubble dynamics analysis.
All authors read and commented on the manuscript. All authors approved the final version of the manuscript.

\section{Competing interests}
The authors declare no competing interests.

\bibliography{1_references}

\end{document}


\section{Bubble dynamics in microgravity environment}
\begin{figure}
	\includegraphics[width=1\textwidth]{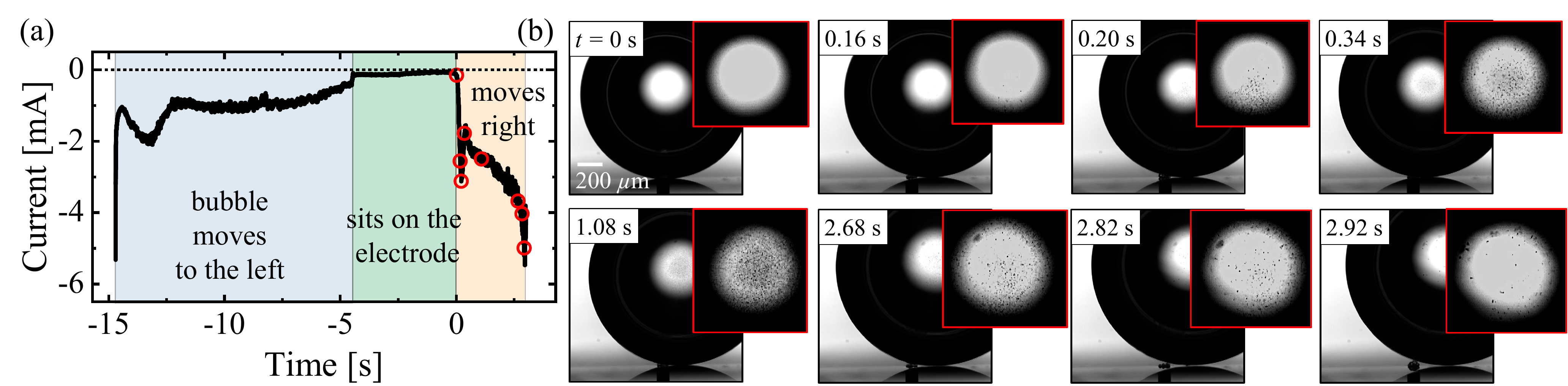}
	\caption{The dynamics of H$_2$ bubble in micro-$g$ achieved during parabolic flights: (a) electric current over the entire evolution cycle; (b) shadowgraphs of the bubble moving laterally at the instants of time marked by the red circles in (a). The insets zoom into the central part of the bubble.
	\label{E:fig1}}
\end{figure}

Figure \ref{E:fig1} demonstrates the evolution of H$_2$ bubble during water electrolysis at $-4$~V (vs Pt wire) in 0.5~mol/L H$_2$SO$_4$ in a micro-$g$ environment, i.e. at greatly eliminated buoyancy, achieved during the parabolic flights (see Bashkatov et al. \cite{bashkatov2021dynamics}).
The electric current in sub-figure (a) reflects the dynamics of the bubble from its nucleation ($t = -14.71$ s) and until detachment ($t$ = 2.94 s). The shadowgraphs in sub-figure (b) depict the bubble position at the time instants marked by red circles in sub-figure (a).

Here, due to a residual gravitation acceleration and its sign variation, the bubble evolution is characterized by the lateral motion shortly after its nucleation. 
During the first part of the evolution, it moves to the left (marked by grey shading in Fig.~\ref{E:fig1}a) relative to the electrode center. Later on, it rolls over the electrode moving to the right while continuing to grow via coalescence with the carpet of microbubbles and diffusion. 
As the bubble rolls over the electrode it blocks most of its active area, minimizing the electric current (marked by green shading in Fig.~\ref{E:fig1}a), hence electrochemical reaction and hydrogen production. The first snapshot ($t=0$) marks the time instant shortly before the bubble releases the electrode. The insets zoom into the central transparent part of the bubble.
From the second snapshot onward, the bubble releases the electrode, drastically increasing the electric current and allowing the formation of a carpet of microbubbles.
The mother bubble continuously coalesces with newly nucleated bubbles (carpet) in the radial direction on a time scale of $\mathcal{O}$($\mu$s). The successive images reveal the internal flow of the microdroplets, emerging from the bottom, presumably at the coalescence spot, and flowing to the top, shortly after coalescence events begin. These droplets can be already seen in the second snapshot ($t$ = 0.16~s). Since the bubble is displaced from the electrode center, the coalescence occurring in the radial direction results in the asymmetrical flow. Upon bubble departure, i.e., when no coalescence occurs, the already injected droplets exhibit minimal movement, drifting slowly due to residual velocity. For more details, we refer the readers to the main manuscript.

\section{Analytical aberration correction}
Aberrations (optical distortions) caused by light refraction at the curved gas-liquid interface of the bubble lead to a significant systematic measurement
deviation\cite{Bilsing2024, gao2021distortion,radner2020field}. As the bubbles are of sub-millimeter size, the correction using an optical element is challenging. However, for object-space telecentric lenses, like the microscope objective used for the measurements here, the aberrations can
be corrected analytically, which has already been done for the flows inside the droplets\cite{kang2004quantitative,minor2007optical,minor2009flow}. In the present study, the entrance pupil of the system is at infinity resulting in chief rays parallel to the optical axis in the object space.  
When back-propagating the rays through the optical system, the intersection of the back-propagated ray and the light sheet (positioned in the middle of the bubble), can be calculated with and without the bubble in the system.

\begin{figure}
	\includegraphics[width=0.8\textwidth]{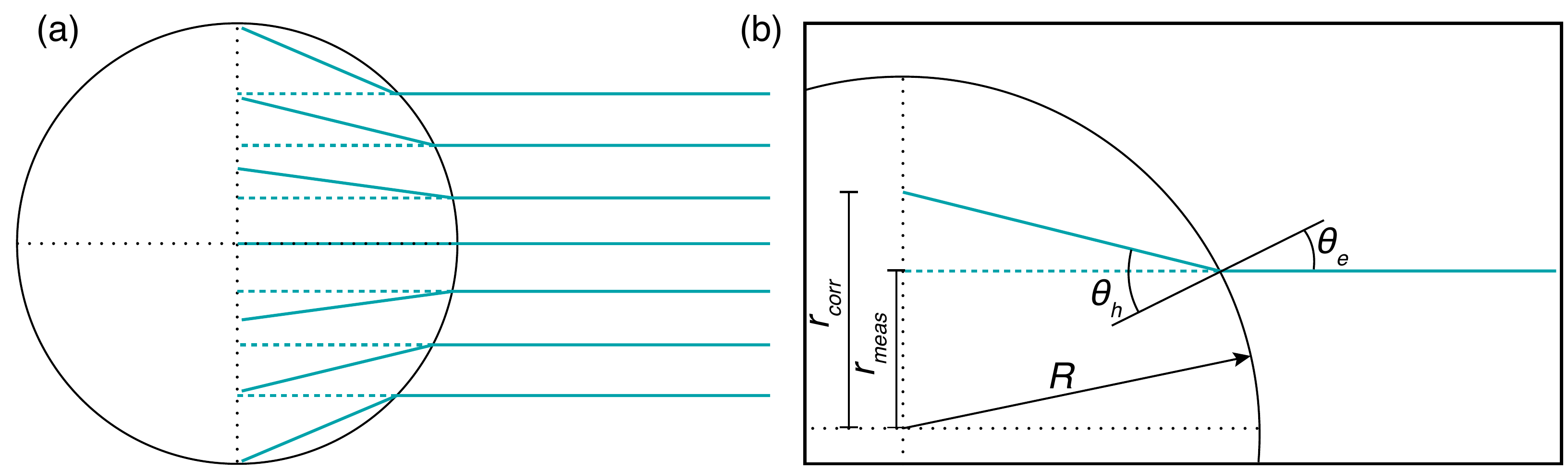}
	\caption{(a) Scheme of the chief rays in an object-space telecentric setup. The microscope objective is on the right side. The solid lines (cyan color) mark the real path of the light, whereas the dashed lines indicate the position in the image plane recorded by the camera. (b) A close view of a single ray passing through the bubble-electrolyte interface with the relevant geometry used in the calculations of the corrected position.
		\label{ESI:fig3}}
\end{figure}

Figure \ref{ESI:fig3} documents a schematic of the chief rays passing through the bubble surface. The solid lines mark the real path of the light scattered at the injected electrolyte droplets, whereas the dashed lines indicate the path recorded by the camera. Fig. \ref{ESI:fig3}b demonstrates a single ray passing through the bubble-electrolyte interface with the relevant geometry used in further calculations of the corrected position for each detected droplet. Since the bubble is assumed to be axisymmetric, the position of the droplet is defined by the radial distance from the bubble center in the plane of the laser sheet. The measured and real (corrected) positions of the injected droplet are therefore defined as $r_{meas}$ and $r_{corr}$. The corrected position for each of the detected  droplets can be calculated as $r_{corr} = r_{meas}+\Delta r$ using

\begin{equation}
    \Delta r = \tan\left[\sin ^{-1} \left[n\cdot \sin \left[\tan^{-1} \left(\frac{r_{meas}}{\sqrt{R^2-r_{meas}^2}} \right) \right]\right] - \tan^{-1} \left(\frac{r_{meas}}{\sqrt{R^2-r_{meas}^2}} \right) \right]\cdot \sqrt{R^2 - r_{meas}^2}, 
\end{equation}
\\
where $n$ is the refractive index of the electrolyte and $R$ is the radius of the bubble.

\begin{figure}\includegraphics[width=0.8\textwidth]{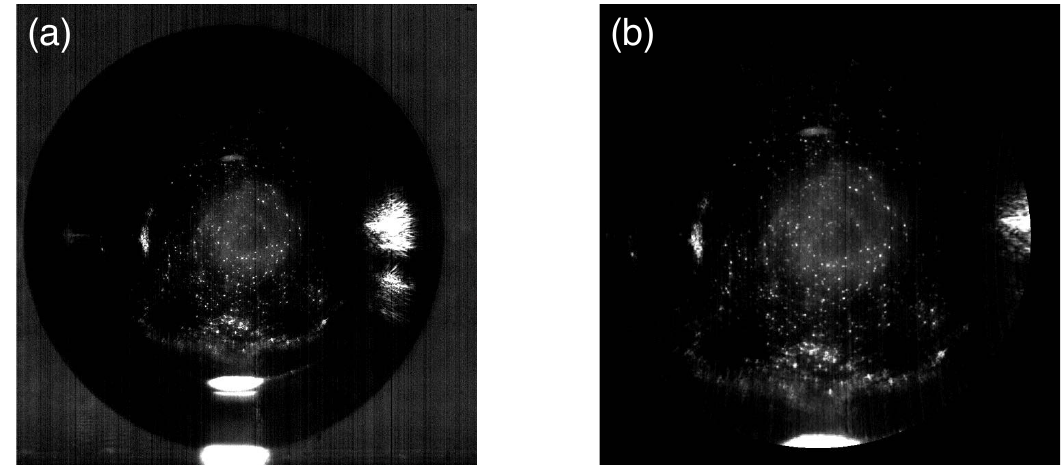}
	\caption{(a) Original snapshot recorded by the camera. (b) Image representing the corrected position of the droplets. The pixels beyond the bubble interface are set to zero (black) in (b). 
		\label{ESI:fig4}}
\end{figure}

Figure \ref{ESI:fig4} represents the effect of the correction with (a) showing a raw snapshot of a H$_2$ bubble and the electrolyte droplets inside; and (b) showing the same snapshot after performing the correction. 
In the raw image, the droplets are concentrated in the central part of the bubble away from the interface. This is due to the curvature of the bubble, which distorts the path of the rays. This is especially visible at the bottom of the bubble. The light scattered by the microbubble carpet (see manuscript) reaches the camera sensor both directly through the electrolyte and through the bubble. The light traveling through the bubble seems to come from a position closer to the center. When the correction is applied, the light traveling through the bubble seems to come from the bubble surface, which implies that the analytical model is correct (Fig. \ref{ESI:fig4}b). 
Note that the model assumes a spherical shape of the droplet as an approximation. Deviations from the spherical shape of the refracting surface caused e.g. by capillary waves or during detachment can potentially be corrected by approaches using adaptive optics \cite{Bilsing2024, gao2021distortion,radner2020field}.

\section{Bubble-carpet coalescence prior departure}
\begin{figure}
	\includegraphics[width=1\textwidth]{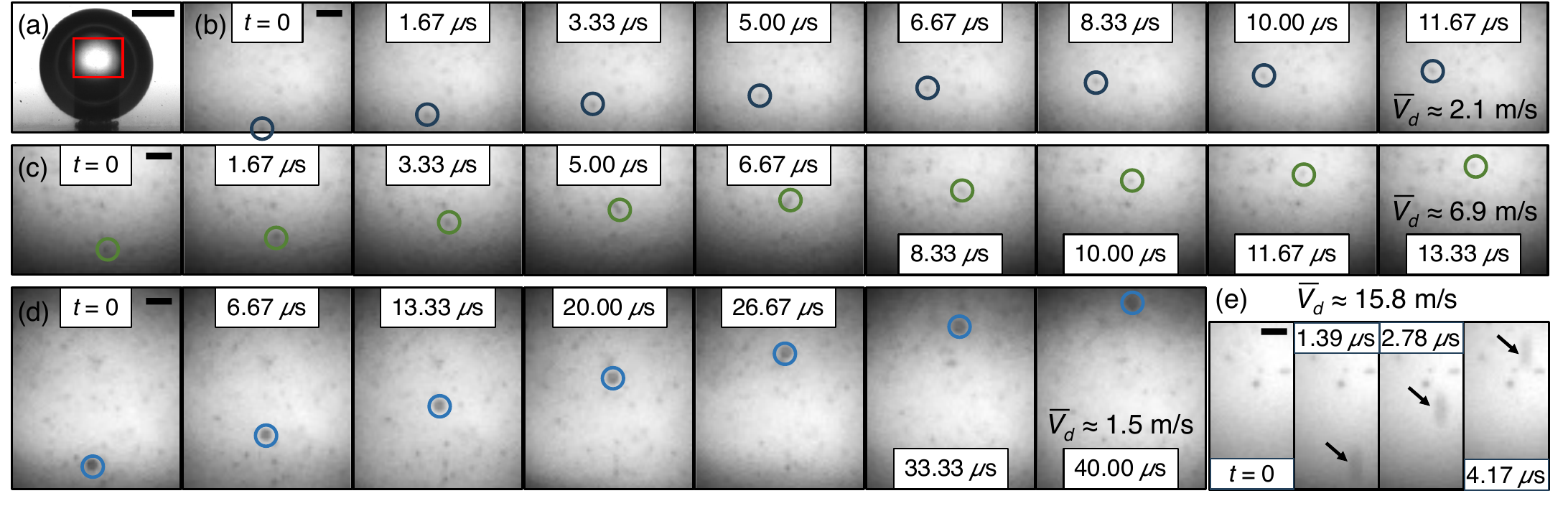}
	\caption{(a) H$_2$ bubble growing at the electrode at $-2.8$~V (vs RHE). (b-e) zoom-in on the central segment of the bubble shown by red rectangular in (a), demonstrating the motion of the injected electrolyte droplets with the velocity $\overline{V}_d$. The image recording was performed at 600~kHz and 720~kHz in (b-d) and (e), respectively. The scale bar is 100~$\mu$m in (a) and 10~$\mu$m in (b-e).
		\label{ESI:fig1}}
\end{figure}

Figure \ref{ESI:fig1}a documents the H$_2$ bubble shortly before its departure in 0.1~mol/L H$_2$SO$_4$ at $-2.8$~V (vs RHE). The bubble sits above the electrode surface at the carpet of microbubbles having thickness $\delta$ and continuously coalesces with it. Figure \ref{ESI:fig1}(b-e) zooms in on the central segment of the bubble marked by red rectangular in (a). The image recording was performed at 600~kHz and 720~kHz in (b-d) and (e), respectively.

The high-speed recording demonstrates that some of the droplets are injected at velocities three orders of magnitude larger than a majority of the droplets, especially during the bubble growth phase, reaching up to 15.8~m/s. These events are rather rare and can be observed more frequently shortly before and at the bubble departure, i.e. when the thickness of the carpet is maximum, about $\delta=16$~$\mu$m and above. The larger $\delta$ will also result in droplets of bigger size.

\section{Stokes's drag: droplet velocity over time upon injection}
The model experiment (see Fig. 4 in the main text) demonstrated that the coalescence of two unequal-sized bubbles is followed by an upward liquid jet, known as the Worthington jet, propagating inside of the merging H$_2$ bubbles. The jet will eventually break into droplet(s) due to the Rayleigh--Plateau instability moving with the velocities of m/s order of magnitude. 

In the Stokes regime ($Re < 1$), the injected droplet would experience drag (frictional) force and decelerate quickly due to viscous drag within the surrounding H$_2$ gas. For reference, the Reynolds number given as $Re = (\rho_{H2}\cdot V_{d}\cdot 2R_d)/\mu_{H2}$ equals to 0.5 with the radius of the droplet $R_d = 5\:\mu$m and velocity of the droplet $V_d=5$ m/s. $\rho_{H2}$ and $\mu_{H2}$ are the density and dynamic viscosity of the hydrogen. We apply Newton's second law to the drag force,

\begin{equation}
\label{eq:F}
    m_{d}\cdot \frac{dV_{d}}{dt} = F_{drag} = -6 \cdot \pi \cdot \mu_{H2}\cdot R_{d} \cdot V_{d},
\end{equation}
\\
where $m_d$ is the mass of the droplet. By integrating Eq. \ref{eq:F}, the velocity of the droplet over time reads

\begin{equation}
\label{eq:ESI-v0}
V_{d}(t) = V_0 \cdot \exp\left(-t/\tau_d
\right), 
\end{equation}
\\
where $\tau_d = \frac{m_d}{6 \cdot \pi \cdot \mu_{H2}\cdot R_{d}}$ and $V_0$ is an initial droplet velocity at the separation from the jet. Typically, $\tau_d \approx 25$~$\mu$s ($R_d = 1$~$\mu$m) and $\tau_d \approx 650$~$\mu$s ($R_d = 5$~$\mu$m), so drag quickly brings the flying drop to a stand-still.

\begin{figure}
	\includegraphics[width=1\textwidth]{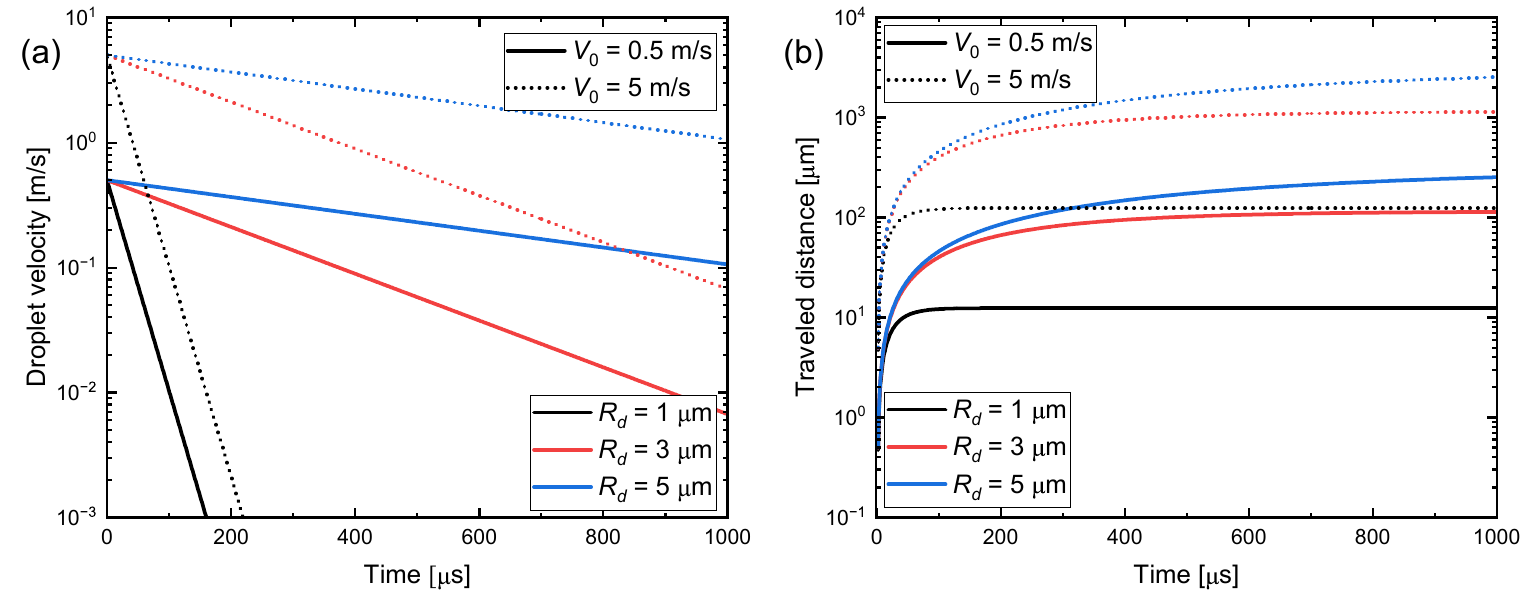}
	\caption{(a) Velocity ($V_d$) and (b) traveled distance ($S_d$) of the injected droplet over time.
		\label{ESI:fig2}}
\end{figure}

Figure \ref{ESI:fig2} demonstrates (a) the velocity $V_d(t)$ and (b) the traveled distance $S_d(t) = \int_{0}^{t} V_d(t') \,dt'$ of the injected droplet over time at three various sizes of $R_d=1$~$\mu$m, $R_d=3$~$\mu$m, and $R_d=5$~$\mu$m, plotted in black, red, and blue colors, respectively and with the initial velocities of $V_0 = 0.5$~m/s and $V_0 = 5$~m/s at $t_0=0$, plotted as solid and dotted lines, respectively. $S_d(t)$ is calculated by integrating the $V_d(t)$ in subfigure (a). Both plots are on a semi-logarithmic scale.

Upon the injection, the droplet velocity thus decays exponentially with time. The droplet with $R_d = 1\:\mu$m (implies $\tau_d \approx 25$~$\mu$s) and with $V_0 = 5$~m/s slows down to the velocity of $10^{-2}$~m/s at $t = 160$~$\mu$s by travelling $S_d = 124$~$\mu$m.  
In comparison, by reducing the initial injection velocity ($V_0$), the droplet will reach the same order of magnitude ($10^{-2}$~m/s) at $t = 100$~$\mu$s by traveling only $S_d = 12$~$\mu$m. By increasing the droplet size from $R_d = 1$~$\mu$m to $R_d = 3$~$\mu$m and $R_d = 5$~$\mu$m, the effect of viscous drag progressively reduces, so that the droplet at $V_0=5$~m/s can cover distances larger than 1 mm before the final position is reached (see subfigure b).

The droplet velocity $V_d$ of $10^{-2}$ mm/s order of magnitude approximately resembles the one found in Fig. 2 (main text) at the bubble bottom in the bubble-carpet system.

While it is a non-trivial to identify the injection time due to continuous bubble-carpet coalescence events start shortly upon bubble nucleation, i.e. after the formation of a single bubble at the electrode surface, we hypothesize that the velocity is resolved away from the injection area at the distance somewhat in between the above-mentioned cases $S_d = 12\:\mu$m and $S_d = 100\:\mu$m. Rather, it is closer to the $S_d = 12\:\mu$m and therefore we tend to assume the initial velocity is close to $V_0 = 0.5$~m/s.

This approximately resembles the velocity of the droplets ($R_d = 1$~$\mu$m) in Fig. 2 (main text) resolved at approx. the same distance away from the injection cross-section at the bubble bottom in the bubble-carpet system.

Besides the initial conditions such as the size ratio of two coalescing bubbles and broken symmetry of the coalescence as in the case of the bubble-carpet system (Figs. 1 and 2 in the main text), the initial velocity of the droplet will depend on at which moment the droplet separates from the jet, i.e. whether it is the first or second droplet as in case of Fig. 4 (see main text). The last separation happened at the retraction of the jet, and therefore with a much smaller velocity, at 1.1 m/s and $R_d=11\:\mu$m for the second droplet vs 7.4 m/s and $R_d=7\:\mu$m for the first droplet. The solution is not trivial, since the first droplets are of smaller size than the last ones, and therefore slows down much faster, however, separating at a much faster initial velocity $V_0$.

In the case of the surface-attached bubble (Fig. 3, main text), the injected droplet slows down upon the injection due to the Stokes's drag and then falls at the bubble-electrode contact area due to gravitational forces, presumably with the terminal velocity. In the Stokes regime ($Re < 1$), the terminal velocity of the droplet can be written

\begin{equation}
    V_t = \frac{2}{9} \frac{R_{d}^2\Delta \rho g}{\mu_{H2}},
\end{equation}
\\
where $\Delta \rho = \rho_{el} - \rho_{H2}$ is the electrolyte-gas density difference and $\mu_{H2}$ the dynamic viscosity of the hydrogen. Given that the second observed droplet has a radius of about $r=4$~$\mu$m, the terminal velocity $V_t = 4.1$ mm/s. 
This value closely correlates with the experimentally estimated velocity of $\overline{V}_d = 6$~mm/s during the period from $t = 1.33$ ms after injection until the droplet reaches the bubble-electrode contact area at $t = 86.67$~ms.

\section{Mechanism of injection: experiment vs simulation}
The direct numerical simulation accurately reproduces the key features and timescales of phenomena such as neck formation, capillary wave propagation, jet formation, and droplet breakup. We then proceed to compare the details of drop ejection between the experiments and simulations presented in the manuscript body in Fig. 4. In the experiments, the first droplet with a radius of \( R_d = 13 \, \mu \text{m} \) is observed at \( t = 250.0 \, \mu \text{s} \) and ejects with a velocity of approximately \( 7.2 \, \text{m/s} \). The second droplet, with a radius of \( R_d = 24 \, \mu \text{m} \), appears at \( t = 341.7 \, \mu \text{s} \) and separates just before the jet starts to retract between \( t = 333.3 \, \mu \text{s} \) and \( t = 341.7 \, \mu \text{s} \), resulting in a much smaller velocity of about \( 1.1 \, \text{m/s} \). In contrast, in the simulations, the first droplet (radius \( R_d = 15 \, \mu \text{m} \)) pinches off at \( t = 260 \, \mu \text{s} \) with a velocity of \( \overline{V}_d = 4.3 \, \text{m/s} \). The second droplet (radius \( R_d = 18 \, \mu \text{m} \)) pinches off at \( t = 285 \, \mu \text{s} \) with a velocity of \( \overline{V}_d = 3.4 \, \text{m/s} \). Additionally, another droplet (radius \( R_d = 3 \, \mu \text{m} \)) was detected at an earlier phase (\( t = 208 \, \mu \text{s} \)) with a much faster velocity of \( \overline{V}_d = 43 \, \text{m/s} \), likely due to numerical artifacts caused by finite mesh resolution. This droplet can be barely seen in Fig.~4 (main manuscript) at $t=208\:\mu$s. The characteristics of the injected droplets are summarised in Table \ref{tab:1}.

\begin{table}[h]
\begin{tabular}{ccccccc}
\Xhline{2\arrayrulewidth}
\multicolumn{1}{c}{\multirow{2}{*}{Droplet {[}\#{]}}} & \multicolumn{2}{c}{\textit{t} {[}$\mu$s{]}} & \multicolumn{2}{c}{\textit{R}$_d$ {[}$\mu$m{]}} & \multicolumn{2}{c}{$\overline{V}_d$ {[}m/s{]}} \\ \cline{2-7} 
\multicolumn{1}{c}{}                                  & experiment     & simulation    & experiment      & simulation      & experiment       & simulation      \\ \hline
0 & - &208 & - & 3 & - &  43 \\ \hline
1 & 250 & 260 & 13 & 15 & 7.2 & 4.3 \\ \hline
2 & 341.7 & 285 & 24  & 18 & 1.1 & 3.4 \\ \Xhline
{2\arrayrulewidth}
\end{tabular}
\caption{Comparison of the droplet injection characteristics between experiment and simulations. $t$ represents the time from the onset of coalescence, $R_d$ and $\overline{V}_d$ the radius and the velocity of the injected droplet, respectively. In the experiment, only two droplets ($\#$1 and $\#$2) are observed, whereas the simulation shows three droplets ($\#$0, $\#$1, and $\#$2). Droplet $\#$0 is likely resolved due to numerical artifacts caused by finite mesh resolution.}
\label{tab:1}
\end{table}

The comparison shows qualitative agreement between experiments and simulations, though quantitative discrepancies remain. Experimental estimates of droplet size and velocity may be slightly inaccurate due to the bubble curvature, which could be analytically corrected for a spherical shape. However, the merged bubble deforms significantly upon droplet injection which complicates precise measurements. We stress that the breakup process is a finite-time singularity \cite{zeff2000singularity}, naturally leading to a qualitative agreement between experiments and simulations. However, the time to break up and the velocity of the ejected drop are sensitive to the discrete nature of simulation equations and experimental noise or measurement technique \cite{eggers2024coalescence}. Additionally, the jet's velocity, which determines the droplet's post-ejection velocity, varies over time\cite{deike2018dynamics,sanjay2021bursting}; small variations in sampling time can thus explain discrepancies. Factors such as an inaccurate gas-liquid viscosity ratio and the sensitivity of velocity to the Ohnesorge number and size ratio also contribute to the differences between experimental and simulation results\cite{walls2015jet}. Given these error sources, the observed discrepancies are reasonable.

\bibliography{2_supp}